\documentclass[preprint,showpacs,preprintnumbers,amsmath,amssymb]{revtex4}

	\usepackage{graphicx,dcolumn,bm}
\begin{document}
\title{Gate Coupling to Nanoscale Electronics}

\author{Sujit S. Datta}
 \affiliation{Department of Physics and Astronomy, University of Pennsylvania, Philadelphia, Pennsylvania 19104, USA\\
Department of Physics, Harvard University, Cambridge, Massachusetts 02138, USA}

\author{Douglas R. Strachan}
 \email{doug.strachan@uky.edu}
\affiliation{Department of Physics and Astronomy, University of Pennsylvania, Philadelphia, Pennsylvania 19104, USA\\
Department of Physics and Astronomy, University of Kentucky, Lexington, Kentucky 40506, USA}

\author{A. T. Charlie Johnson}
 \email{cjohnson@physics.upenn.edu}
\affiliation{Department of Physics and Astronomy, University of Pennsylvania, Philadelphia, Pennsylvania 19104, USA}

\date{\today}

\begin{abstract}
The realization of single-molecule electronic devices, in which a nanometer-scale molecule is connected to macroscopic leads, requires the reproducible production of highly ordered nanoscale gaps in which a molecule of interest is electrostatically coupled to nearby gate electrodes. Understanding how the molecule-gate coupling depends on key parameters is crucial for the development of high-performance devices. Here we directly address this, presenting two- and three-dimensional finite-element electrostatic simulations of the electrode geometries formed using emerging fabrication techniques. We quantify the gate coupling intrinsic to these devices, exploring the roles of parameters believed to be relevant to such devices. These include the thickness and nature of the dielectric used, and the gate screening due to different device geometries. On the single-molecule ($\sim$1nm) scale, we find that device geometry plays a greater role in the gate coupling than the dielectric constant or the thickness of the insulator. Compared to the typical uniform nanogap electrode geometry envisioned, we find that non-uniform tapered electrodes yield a significant three orders of magnitude improvement in gate coupling.  We also find that in the tapered geometry the polarizability of a molecular channel works to enhance the gate coupling. 
\end{abstract}
\maketitle
\section{\label{sec:level1}Introduction}
Three-terminal electronic devices constructed on the single-molecule scale ($\sim$1 nm) have received intense recent interest and hold the potential to be the smallest possible transistors [1-3].  These nano-scale devices are expected to share a number of similarities with their larger predecessors, the ubiquitous semiconducting field-effect transistor [4] and the single electron transistor constructed from two-dimensional electron systems [5, 6] and small ($\sim$100 nm) metallic islands [7].  The architecture of molecular-scale devices is similar to that of their predecessors, with a single molecule coupled to source and drain electrodes acting as the conduction channel. The value of the source-drain current $I_{sd}$ is then modulated through the use of a nearby gate electrode controlling the potential of the channel.  So far, three terminal devices have been reproducibly constructed with channel lengths larger than $\sim$20 nm using nano-scale materials such as carbon nanotubes [8], nanowires [9], graphene segments [10], molecular films [11], and large ($>$20 nm) colloidal particles [12]. 

To reduce the dimensions of three-terminal devices to the single-molecule scale (well below 20 nm) requires that a number of size-scaling issues be addressed.  In particular, it is well known that for ultra-short channels it becomes increasingly difficult to influence the channel electrical conduction via the modulated potential of the gate electrode [13]. Diminished gate coupling at the nano-scale has already become an issue in the construction of carbon nanotube transistors [14], molecular thin-film transistors [15], and semi-conducting nanowire devices [16].  To address this issue there has been interest in using ultra-thin high-$\kappa$ dielectrics to increase the gate coupling as channel dimensions shrink towards 20 nm [17, 18].  Yet on the much smaller molecular-scale (of order 1 nm), it is still not clear how to most effectively couple the gate electrode to the channel [19].  It has also recently become apparent that the possibility of unplanned tunnel and/or capacitive coupling to metal particles or other contaminants is a serious concern for these devices [20-23].

Here we present detailed investigations of the geometrical effects for gate coupling to single-molecule-scaled devices using finite-element simulations of the electrostatics.  We find that the precise geometry of the source-drain electrodes has a tremendous effect on the ability to couple the gate to the molecular-scale devices.  In fact, on the single-molecule (1 nm) scale, the geometry plays a greater role in the gate coupling than the dielectric constant or the thickness of the insulator.  Compared to a uniform nanogap electrode geometry that is typically envisioned for molecular-scale devices, we find that non-uniform tapered electrodes [24] yield an improvement in gate coupling by approximately three orders of magnitude.  We also find that in the tapered geometry the polarizability of a molecular channel works to enhance the gate coupling, whereas for the typical uniform nanogap geometry the effect is the opposite.

\section{\label{sec:level1}Model and Method of Calculation}
Figure 1a shows the typical model envisioned [1-3] of a three-terminal molecular-scale electronic device.  The general expectation is that a small molecular-scale object, such as a single molecule or a nanoparticle, could be placed onto the nanogap of a metallic electrode and would thus provide an electrical conducting pathway between the source and drain.  As shown, the source and drain are electrically insulated from the nearby gate by a dielectric layer with the molecule or particle schematically represented as a sphere located at the top of the electrode.  A cross-section is shown in Figure 1b with the critical dimensions of the device labeled: $t$ is the thickness of the electrodes,  $L_{ox}$ and $\kappa$ are the thickness and dielectric constant of the dielectric, respectively, and $s$ is the width of the nanogap. We assume that the bottom of the molecular-scale object coincides with the top of the source and drain -- a reasonable assumption if the object is to bridge both electrodes [25-28].  The light-blue lines in Figure 1b are equipotentials obtained from a finite-element simulation where the gate electrode is charged to a positive (depicted as white) potential while the source and drain are approximately grounded (depicted as black).  

We performed the electrostatic simulations presented in this paper using the \texttt{COMSOL Multiphysics 3.3a} finite-element simulation package.  This solves for relevant electrostatic quantities such as the electrostatic potential $V$ throughout a simulation geometry containing subdomains of specified materials properties and faces of specified boundary conditions using quadratic Lagrange elements.  We  used a stationary linear iterative (Generalized Minimum Residual) solver working with the algebraic multigrid preconditioner appropriate for the scalar elliptic partial differential equations that characterize this system.  The simulation geometry is a vacuum box, with zero charge/symmetry boundary conditions (i.e. $\mathbf n\cdot \mathbf D=0$, where $\mathbf n$ is a vector normal to the face and $\mathbf D$ is the electric displacement field) on all faces except for that coincident with the back-gate, which is an equipotential of voltage $V_{g}$.  The devices we model constitute three-dimensional subdomains within this simulation geometry, with the electrodes taken to have dielectric constant $\kappa=10^{6}$ and voltage $V=0$; thus, they are effectively grounded equipotentials.  We verified that simulation geometry was large enough that small-size boundary effects did not affect the results presented herein.

Throughout the simulations we assume that small-size effects do not significantly alter material parameters - such as the dielectric constants - from their bulk values.  This is supported by recent work with sub-10nm thick dielectrics [18] which demonstrates a $\kappa\sim20$ -- a value well within the range investigated here.  The bulk values we use for the electrodes also amount to the assumption that the screening at the metallic surfaces occurs abruptly over an infinitesimally thin skin depth.  This assumption is justified since the electrostatic screening length at metallic surfaces (e.g., for Cu and Au) is approximately 2$\AA$ [29]. This length scale is at least an order of magnitude smaller than any pertinent physical dimension of the electrodes investigated here. 

\section{\label{sec:level1}Results and Discussion}
\subsection{\label{sec:level2}Electrodes with uniform cross-section}
A voltage $V_{g}$ applied to the gate causes the voltage within the nanogap region to swing towards this potential.  The ability for the gate electrode to influence the electrical conduction of the device depends on how much of the applied gate voltage reaches up to the molecular-scale object near the top of the electrodes.  This can be appreciated by considering the geometry of 6 nm thick source and drain electrodes containing a 2 nm sized nanogap -- small enough to be bridged by typical molecular-scale objects [25-28] -- and situated on top of a 3 nm thick dielectric.  We will focus on the potential $V(z)$ along the vertical center-line of the nanogap and define the zero ($z=0$) at the top of the electrodes where the crossed dashed lines intersect in Figure 1b at the base of the molecule.  

Figure 1c shows the normalized potential $V(z)/V_{g}$ as a function of position for various dielectric materials.  Considering the case of a dielectric with $\kappa=1$ (the lowest curve in Figure 1c), the potential $V(0)$ that reaches the molecular-scale object is reduced from $V_{g}$ by a factor of $10^{5}$. We term this the gate coupling ratio, $\Gamma\equiv V(0)/V_{g}$.  This $10^{5}$ reduction in the potential results from electrostatic screening by the electrodes: this view is supported by the functional behavior of $V(z)$ inside the nanogap, which goes as $e^{-\pi z/s}$ (see for example the dashed line in Figure 1c), where $s$ is the 2 nm width.  The exponential form is the lowest order behavior of $V(z)/V_{g}=(4/\pi)\sum_{n}{\mathrm{exp}(-n\pi z/s)~\mathrm{sin}(n\pi/2)\cdot n^{-1}}$ with $n$ odd, the potential for the scenario where a gate electrode is placed adjacent ($L_{ox}=0$) to two grounded conducting slabs separated by $s$ [30]. Here we use the $n=1$ Fourier component, which is an excellent approximation to exponential decay within the nanogap region.

The consequences of this $10^{5}$ reduction can be understood by considering the voltage modulations required to influence conduction through the molecular-scale object spanning the nanogap.  That is, at room temperature charge carriers in the device will experience roughly 25 mV random variations due to thermal fluctuations, where $k_{B}T\approx eV$ ($k_{B}$ is Boltzmann's constant, $T$ is the local sample temperature, and $e$ is the elementary charge).  At a minimum, the energy modulations at the molecular-scale object due to the applied gate potential will need to be larger than these 25 mV thermal fluctuations in order to control the device behavior.  This requires that \textit{at least 2500 volts} be applied to the gate electrode across the 3 nm thick dielectric: a scenario which would be impossible to sustain against an immediate breakdown of the dielectric [31-35].  Clearly, this sort of transistor would not function with current state of the art dielectric materials. 

To improve the gate coupling, one could utilize a high-$\kappa$ dielectric, as is commonly suggested [18, 35, 36].  The use of a high-$\kappa$ dielectric with $\kappa=80$ is shown in Figure 1c and results in a modest (factor of $\sim$2) improvement of the coupling.  Figure 1d shows the relatively insignificant variation of the gate coupling $\Gamma$ with dielectric constant for nanogaps of 2 nm and 5 nm over this range.  Compared to the orders of magnitude effects due to screening, the dielectric has a relatively modest affect on gate coupling for molecular-scale devices.  

Another parameter that is thought to strongly affect gate coupling is the dielectric thickness $L_{ox}$, with effort taken by some groups to utilize ultra-thin ($\sim$3 nm thick) oxide layers as the dielectric in molecular electronic devices [18, 36].  Yet, as with the dielectric constant, we find that $L_{ox}$ has a relatively small influence on the gate coupling for molecular-scale devices.  For example, by varying $L_{ox}$ from 1 nm to 50 nm in Figure 2a for a nanogap (with $s=2$ nm and $t=6$ nm) we find that $V(z)/V_{g}$ varies by $\sim10^{1.5}$. This is significantly less than the $10^{5}$ reduction due to screening by the electrodes.  Furthermore, we find that as $L_{ox}$ is decreased the gate coupling plateaus to a constant value (Figure 2b), whereas the deleterious effects of dielectric breakdown should become increasingly problematic. Thus, there is little benefit to decreasing the dielectric thickness of a nano-electronic three terminal device when screening by the electrodes is significant.

The slight dependence of gate coupling on dielectric constant and thickness is in sharp contrast to the dramatic influence of the electrode geometry itself.  For instance, increasing the width of the nanogap from 2 nm to 5nm (keeping all other properties constant) results in a $10^{3}$ increase in $\Gamma$, as shown in Figure 1c. Other geometrical aspects of the leads also have dramatic effects on the gate coupling.  For example, the thickness of the leads plays an important role in the gate coupling, as seen in Figure 3.  Figure 3a shows $V(z)/V_{g}$ for 2 nm nanogaps on a 3 nm thick dielectric (with $\kappa=4$).  A decrease in the thickness of the leads from 9 nm to 3 nm shows a substantial $10^{4}$ increase in gate coupling. 

This $10^{4}$ increase in gate coupling for thinner leads can be understood by the reduction in screening; \textit{i.e.}, the ratio of the screening for 3 nm and 9 nm thick leads is approximately $e^{-\pi(3nm)/(2nm)}/e^{-\pi(9nm)/(2nm)}\sim10^{4}$.  This large variation in gate coupling with lead thickness in the 3 nm to 9 nm range is pertinent to the fabrication of molecular-scale devices since high-quality continuous metallic electrodes thinner than 6 nm are difficult to deposit.  For this reason, reports of evaporated or deposited nanogaps (utilizing shadow-mask techniques) are generally much thicker than 6 nm [37, 38].

Lead thickness has a significant effect on three terminal devices when channel lengths are less than $\sim$10 nm.  For example, the lead thickness still plays a role in Figure 3b for the slightly wider ($s=5$ nm) nanogaps, which show a $10^{2}$ variation in gate coupling for lead thickness varying from $t=3$ nm to $t=9$ nm.  Figure 3c demonstrates for 6 nm thick electrodes that the gate coupling for a nanogap is strongly diminished when the width is less than $s=10$ nm. 

\subsection{\label{sec:level2}Electrodes with non-uniform cross-section}
Having demonstrated the importance of the electrode geometry for gate coupling, we now consider the effects of nanogaps that do not have a uniform two-dimensional cross-section.  One such structure that we have recently fabricated using feedback controlled electromigration (FCE) [39] contains tapered nanogap electrodes, with edges formed along the lattice crystal planes [24]. A schematic of this structure is shown in Figure 4a, where the source and drain electrodes are angled down towards the nanogap with an angle $\theta=120^\circ$, as dictated by the crystal structure of the lattice.  In the simulations we utilize electrodes with 2 nm flat ends, as in Figure 4a, which approximate the natural rounding that can occur for experimentally-fabricated nanogaps [23].

Results of three-dimensional (3D) simulations for tapered electrodes protruding from electrodes with uniform cross section are shown in Figure 4b for several lead thicknesses.  A striking result of the tapered electrodes is that $V(z)/V_{g}$ has a minimum and increases as a function of $z$ just below the top surface of the source and drain electrodes.  This means that there is a substantially greater gate coupling for tapered electrodes ($\Gamma_{3D}$) compared to nanogaps with uniform 2D cross section ($\Gamma_{2D}$), as seen in Figure 4c.  The improvement in gate coupling gained by tapered electrodes is most pronounced for narrow nanogaps with thick leads.  For the typical nanogaps investigated in molecular electronics ($\sim$2 nm wide, $>$6 nm thick), the coupling increases by at least a factor of $5\times10^{2}$ for tapered electrodes over the uniform 2D electrodes.  

Another important effect of the tapered electrodes is that the gate coupling variations due to lead thickness are greatly diminished.  At the location of the molecule ($z=0$ in Figure 4b), the gate coupling varies by only a factor of $\sim$8 for lead thicknesses between 3 and 9 nm.  This is a significantly smaller variation than the $10^{4}$ changes in coupling for the uniform 2D nanogaps shown in Figure 3a.  

Having found that gate coupling is significantly enhanced if the source and drain electrodes are tapered as opposed to flat, we now investigate the degree to which the electrodes need to be tapered.  Figure 5a shows $V(z)/V_{g}$ for nanogaps (with $s=2$ nm, $t=3$ nm, $L_{ox}=3$ nm, and $\kappa=4$) having electrode angle $\theta$ varying from $0^\circ$ to $180^\circ$, as defined in Figure 4a. The gate coupling for the corresponding angles is shown Figure 5b.  It is clear that there is a dramatic improvement in gate coupling as the electrode angle is decreased from the $180^\circ$ flat surface towards $0^\circ$.  The majority of this improvement occurs in the vicinity of $180^\circ$, with a factor of 10 improvement in going from $180^\circ$ to $120^\circ$.  This suggests that realizing significant improvements of gate coupling does not require nanowire-type contacts (\textit{i.e.}, with $\theta=0^\circ$).  Thus, utilizing the tapered contacts that occur naturally from the \{111\} planes of FCC lattices [24] should result in a significant improvement in gate coupling over uniform 2D ($\theta=180^\circ$) nanogaps.       

As with uniform 2D nanogaps, we find that gate coupling between tapered electrodes is only weakly influenced by the value of the dielectric constant of the insulator.  This is seen in the $V(z)/V_{g}$ plotted in Figure 6 for nanogaps (with $t=3$ nm, $L_{ox}=3$ nm, $\theta=120^\circ$ and two different values of $s$) having various dielectric constants.  In the range of $\kappa$ from 1 to 80 there is less than a factor of two increase in gate coupling for the higher dielectric constants, which is considerably less than the nearly two orders of magnitude reduction due to screening effects. 

The situation is considerably different for the $L_{ox}$ of tapered nanogap electrodes.  In this case the dielectric thickness plays a significant role compared to the screening, as seen in Figure 7.  Here we see that decreasing $L_{ox}$ gives rise to a similar factor of 10 improvement to the gate coupling as for the uniform 2D nanogaps (compared to Figure 2a).  However, in the case of tapered electrodes, the improvement due to $L_{ox}$ is comparable to the magnitude of the screening effect ($\sim10^{1.5}$).  This makes the dielectric thickness an important parameter in determining the gate coupling for tapered nanogap electrodes.

\subsection{\label{sec:level2}Effect of Molecular Polarizability}
So far, we have investigated the electrostatics in the vicinity of nanogaps without consideration of the materials properties of the molecular-scale object that constitutes the channel.  This has the effect of neglecting the induced polarization of the molecule and its influence on the electrostatics.  The effect of polarizability can be estimated by incorporating a spherical metallic particle having $\kappa=10^{6}$ at the top of the nanogap, as schematically represented in Figures 1a and 4a.  The resulting $V(z)/V_{g}$ plots for the 2D (Figure 8a) and tapered (Figure 8b) nanogaps with metallic particles show only a slight modification when a metallic particle is placed onto the gap -- indicating that our initial assumptions of a non-polarizable molecule do not alter our previous conclusions.  

A close inspection of the simulations in Figure 8 reveals that there is an additional effect that can result in an increase in coupling for tapered nanogap electrodes.  Throughout most of the space below the location of the molecule ($z<0$) there is a negligible variation in the potential for all the nanogap geometries we investigated, as seen in Figures 8a and 8b.  However, at the location of the molecule the potential becomes constant due to the polarizability.  In the 2D case (Figure 8a) this deviation results in a factor of $\sim$2 decrease in gate coupling compared to the case without the molecule, as determined earlier in this paper.  This deviation is similar to the result for the widely spaced tapered 5 nm nanogap electrodes, shown in Figure 8b.  The situation for the closely spaced 2 nm tapered electrodes in Figure 8b, on the other hand, exhibits the complete opposite behavior, with roughly a factor of two \textit{increase} in coupling due to molecular polarizability.  

These deviations can be approximately understood as a result of the ``flattening out" of the potential profile in the vicinity of the molecule ($z=0$). For example, a potential profile with a negative slope at this location will experience a \textit{decrease} in coupling compared to the case without the molecule.  Likewise, a positive slope of the potential at $z=0$ will lead to an increase in coupling due to the introduction of a polarizable species.  Focusing on the first spatial derivative of the potential at the location of the molecule ($z=0$) indicates that this is positive for closely spaced tapered electrodes (Figure 8c).  As the angle of the electrodes is decreased (i.e. as the electrodes become more tapered) the value of the derivative of the potential is larger at the location of the molecule, resulting in increased gate coupling.

\subsection{\label{sec:level2}Molecular Trapping and Tunneling to Tapered Electrodes}
One potential advantage of tapered electrodes over uniform 2D electrodes is that they may make single molecule devices more easily achievable.  In particular, the reduced cross-sectional surface area of the tapered nanogap would make it much more likely that the electrodes were bridged with either a {\it single} molecule or a small number of molecules.  This is in contrast to uniform 2D electrodes which enable many parallel molecular conduction pathways.  Though the reduced cross-sectional surface area of tapered electrodes may reduce the probability for constructing a molecular bridge, a recent report of two-terminal single molecule devices constructed with a scanning tunneling microscope (STM) tip suggests that molecules can in fact reproducibly bridge tapered electrodes [40].

A complete treatment of the geometry dependence of tunneling between the source/drain electrodes and a molecular conduction channel would be an interesting complement to the present work, and is outside the scope of this paper. We do not expect the tunneling from molecules to tapered electrodes should be significantly reduced in comparison to uniform 2D electrodes: because vacuum tunneling probabilities typically decay exponentially on a length scale of several $\AA$ [41], the vast majority of the tunneling occurs within several $\AA$ of the molecule located near the tip of the tapered electrode. Thus, the angle of the taper in Figure 4a should have very little effect on the net tunneling current, and device conductance should be dominated by the electrode region closest to the nanogap. In fact, reported detailed calculations [42] indicate that electrodes which provide {\it fewer} possible sites for contact to the molecular conduction channel (as would be the case for tapered electrodes) might even have slightly {\it increased} electronic coupling to the channel.        
    
\subsection{\label{sec:level2}Implications for Molecular-Scale Devices}
The screening effects that we have quantified here have a number of implications for molecular-scale device performance.  First, the increased coupling that could be obtained through tapered contacts would significantly reduce the gate voltage required to modulate the conductance channel of a device.  Based on our simulations we could expect to obtain a gate coupling of $\sim1/50$ for observed tapered contacts [24], where the thickness of the nanogap electrode pair is $t=3$nm and the electrode angle is $\theta=120^\circ$.  This would mean that at room temperature one could achieve gate coupling effects that could successfully compete against thermal fluctuations for $V_{g}\sim1$V -- much less than the 2500 volts required for uniform cross-section nanogaps.  Though this voltage is small enough to resist breakdown of typical dielectrics over the required size scales, it is likely that even better gate coupling will be required for future molecular-scale electronics.  To enable this would require the ability for molecules to be pulled down towards the interface of the metal/dielectric interface in order to significantly reduce the screening by the electrodes.

Another important implication is that, in the case of significant electrode screening, the gate electrode could much more easily couple to residual parasitic conducting pathways resting on the surface of the dielectric, as opposed to the molecules that would ideally lay at the tops of the nanogap electrodes [3].  Such artifacts could include residual metallic clusters resulting from electromigration [20, 21] or islands that result from the evaporation of the leads [23].  In some cases these metallic artifacts may provide systems with interesting novel phenomena and could mimic some of the properties of single molecule electronics [20-22].  However, in cases where the intended device is a single molecule transistor these artifacts could seriously interfere with the intended device performance.

\section{\label{sec:level1}Conclusions}
Nanogap electrode geometries are emerging as potentially useful candidates in constructing molecular-scale electronic three-terminal devices. To better understand how the molecule-gate coupling depends on key parameters, we have investigated the electrostatics of these devices. The screening effect due to electrode geometry is found to be the most significant parameter in determining gate coupling to the molecular-scale channel, in comparison to the properties of the dielectric. Compared to the uniform nanogap electrode geometry that is typically envisioned for molecular-scale devices, we find that non-uniform tapered electrodes [24] could obtain approximately three orders of magnitude improvement in gate coupling. An additional enhancement in the gate coupling is also found to occur in the tapered geometry due to the polarizability of the molecular channel -- this is in contrast to a uniform nanogap geometry, where the polarizability of the molecular device instead causes the gate coupling to decrease. 

\begin{acknowledgments}
This work was supported by National Science Foundation (NSF) Grant DMR-0805136, the University of Pennsylvania Nano/Bio Interface Center through NSF NSEC DMR-0425780, and the Intelligence Community Postdoctoral Fellowship Program.
\end{acknowledgments}

\newpage

\begin{figure}
\begin{flushright}
\includegraphics[width=5.5in]{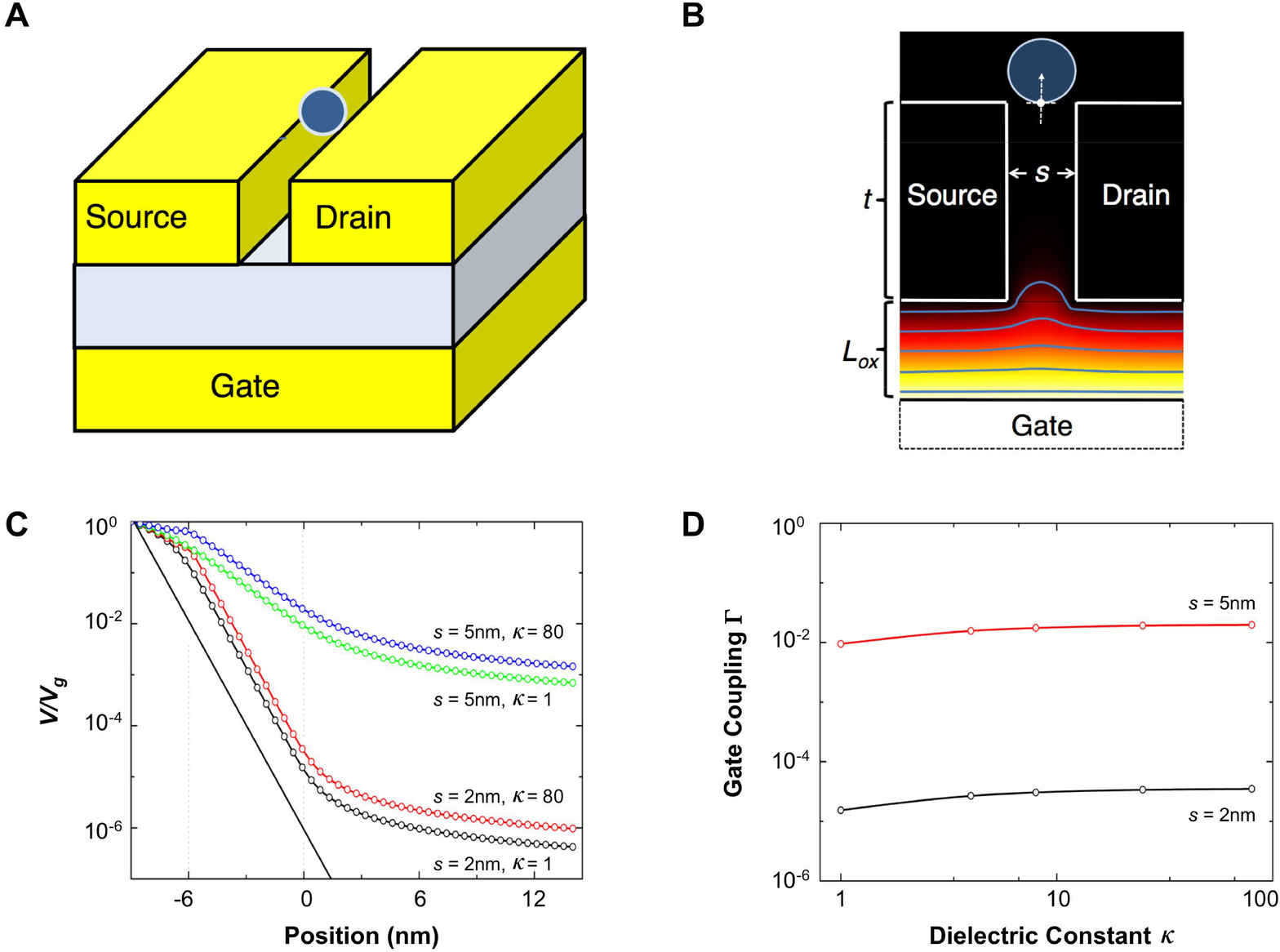}
\caption{(Color Online) {\bf Electrodes with uniform cross-section -- role of the dielectric constant. {\bf (A)}} Schematic of a typical three-terminal molecular-scale electronic device; {\bf (B)} Vertical cross-section through the device shown in A, with calculated equipotentials (light blue) and electrostatic potential (heat map); {\bf (C)} Normalized potential for the two-dimensional device shown in A with $L_{ox}=3$nm, $t=6$nm, including the analytic form of the potential (thick black line, see section 3.1 of main text for details). Throughout this paper, the left dotted line indicates the interface between the gate dielectric and the electrodes, while the right dotted line indicates the top of the electrodes; {\bf (D)} Variation of the gate coupling with dielectric constant for nanogaps of two different sizes, for the two-dimensional device shown in A with $L_{ox}=3$nm, $t=6$nm.}
\end{flushright}
\end{figure}
\newpage

\begin{figure}
\begin{flushright}
\includegraphics[width=5.5in]{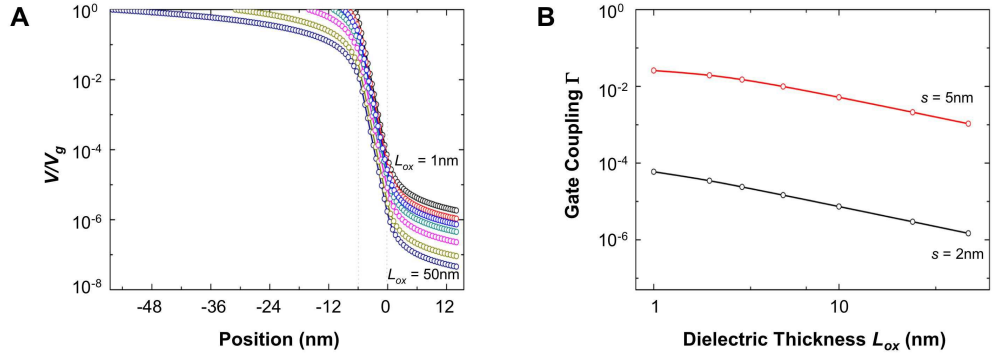}
\caption{(Color Online) {\bf Electrodes with uniform cross-section -- role of the dielectric thickness. {\bf (A)}} Normalized potential for the two-dimensional device shown in Figure 1A with $t=6$nm, $s=2$nm, $\kappa=4$, values of $L_{ox}$ from top to bottom are 1, 2, 3, 5, 10, 25, and 50nm; {\bf (B)} Gate coupling as a function of dielectric thickness for the two-dimensional device shown in Figure 1A with $t=6$nm, $\kappa=4$.}
\end{flushright}
\end{figure}
\newpage

\begin{figure}
\begin{flushright}
\includegraphics[width=5.5in]{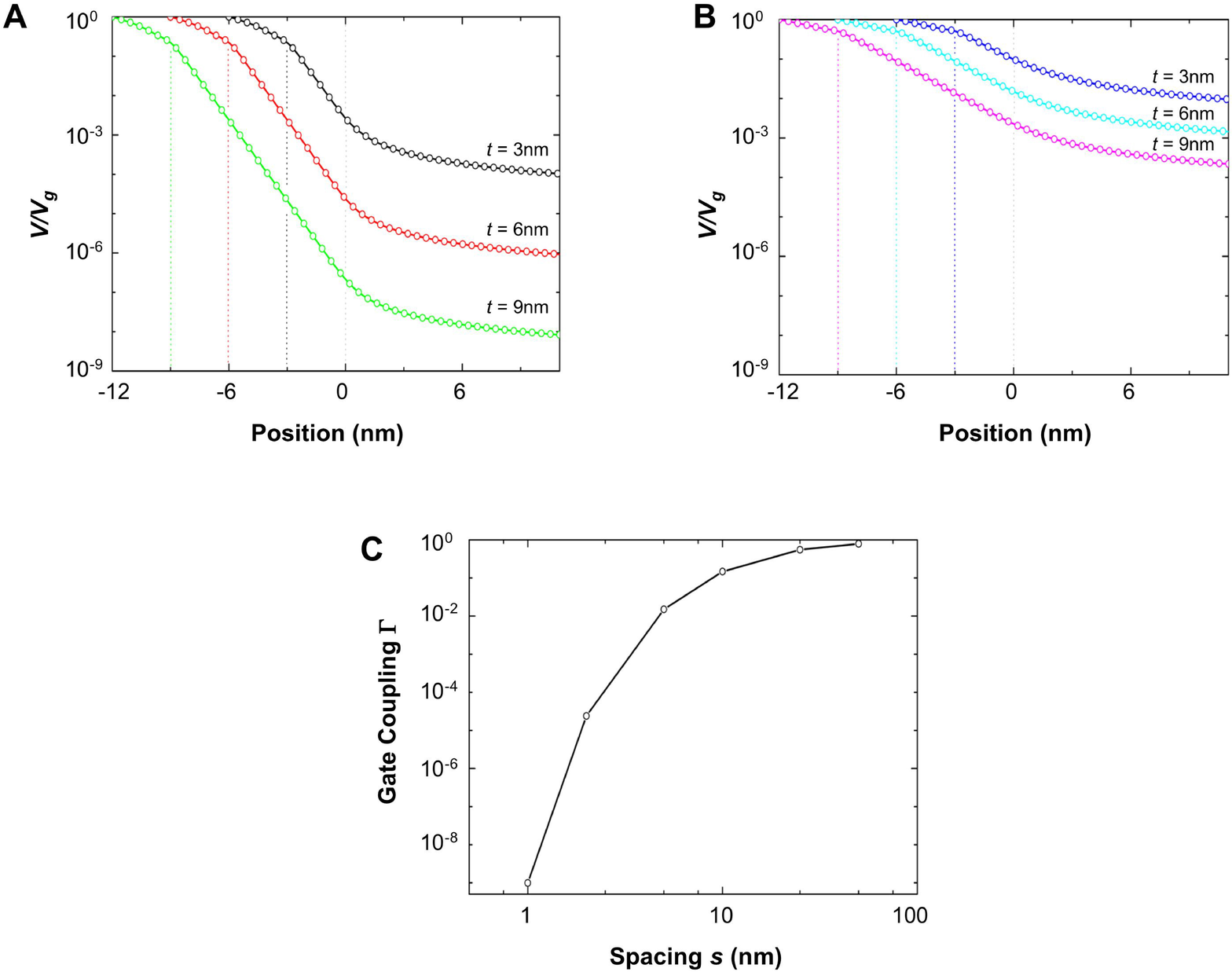}
\caption{(Color Online) {\bf Electrodes with uniform cross-section -- roles of the electrode thickness and spacing. {\bf (A)}} Normalized potential for the two-dimensional device shown in Figure 1A with $L_{ox}=3$nm, $s=2$nm, $\kappa=4$; {\bf (B)} Normalized potential for the two-dimensional device shown in Figure 1A with $L_{ox}=3$nm, $s=5$nm, $\kappa=4$; {\bf (C)} Gate coupling as a function of nanogap size for the two-dimensional device shown in Figure 1A with $L_{ox}=3$nm, $t=6$nm, $\kappa=4$.}
\end{flushright}
\end{figure}
\newpage

\begin{figure}
\begin{flushright}
\includegraphics[width=5.5in]{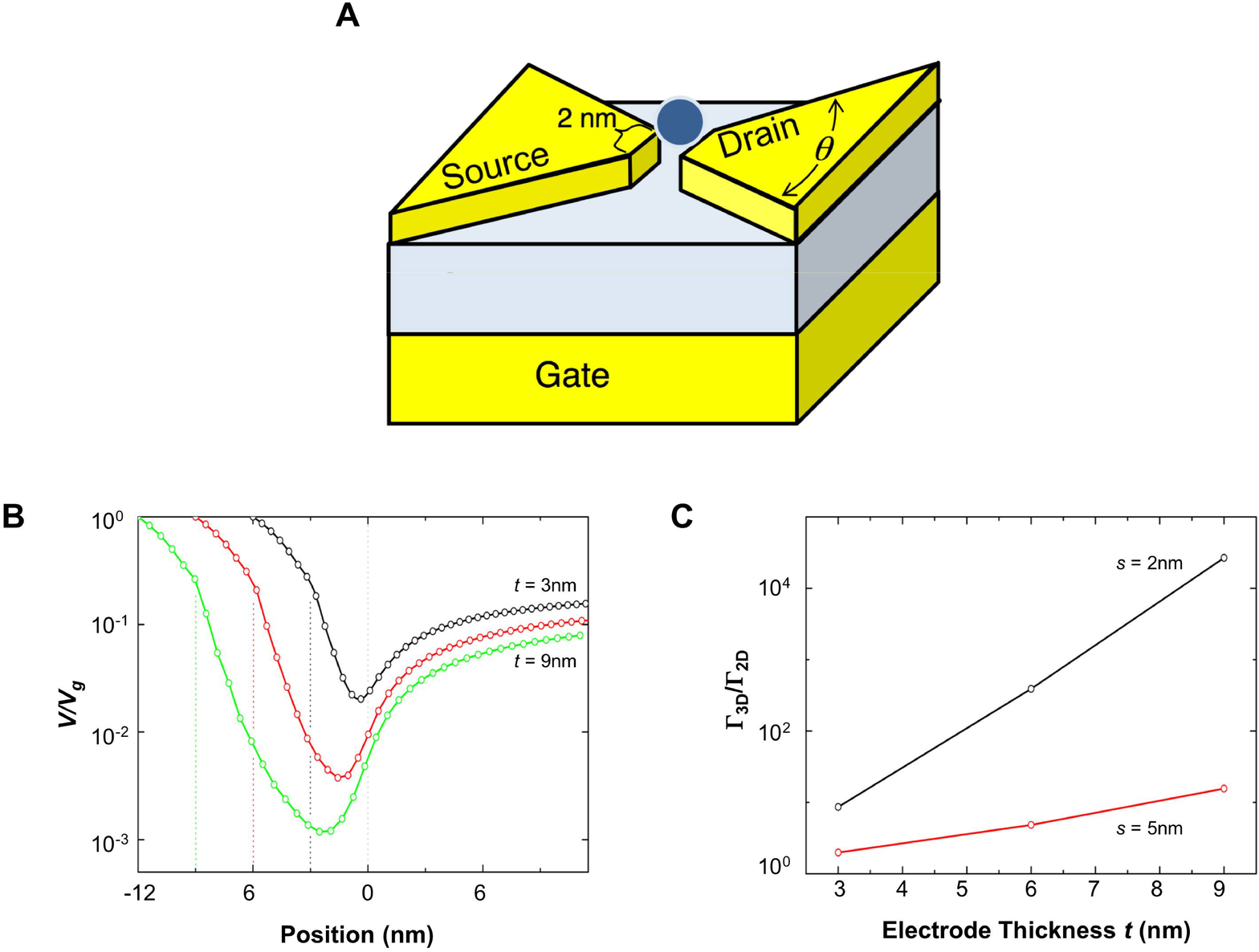}
\caption{(Color Online) {\bf Electrodes with non-uniform cross-section -- role of the electrode thickness. {\bf (A)}} Schematic of a typical three-terminal molecular-scale electronic device fabricated using feedback controlled electromigration; {\bf (B)} Normalized potential for the three-dimensional device shown in A with $L_{ox}=3$nm, $s=2$nm, $\kappa=4$, $\theta=120^\circ$, values of $t$ from top to bottom are 3, 6, 9nm; {\bf (C)} Ratio between gate coupling factor $\Gamma$ for 3D electrodes with $\theta=120^\circ$ and for 2D electrodes, both with $L_{ox}=3$nm, $\kappa=4$.}
\end{flushright}
\end{figure}
\newpage

\begin{figure}
\begin{flushright}
\includegraphics[width=5.5in]{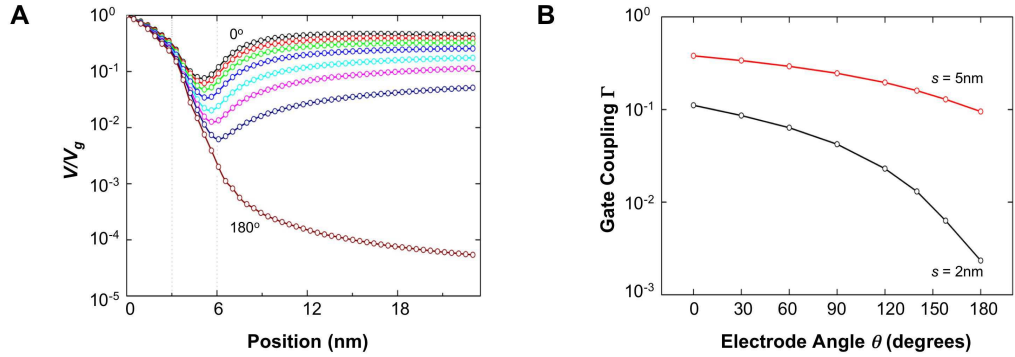}
\caption{(Color Online) {\bf Electrodes with non-uniform cross-section -- role of the electrode angle. {\bf (A)}} Normalized potential for the three-dimensional device shown in Figure 4A with $L_{ox}=3$nm, $t=3$nm, $s=2$nm, $\kappa=4$, values of $\theta$ from top to bottom are 0$^\circ$, 30$^\circ$, 60$^\circ$, 90$^\circ$, 120$^\circ$, 140$^\circ$, 160$^\circ$, and 180$^\circ$; {\bf (B)} Gate coupling as a function of electrode angle for the three-dimensional device shown in Figure 4A with $L_{ox}=3$nm, $t=3$nm, $\kappa=4$.}
\end{flushright}
\end{figure}
\newpage

\begin{figure}
\begin{flushright}
\includegraphics[width=5.5in]{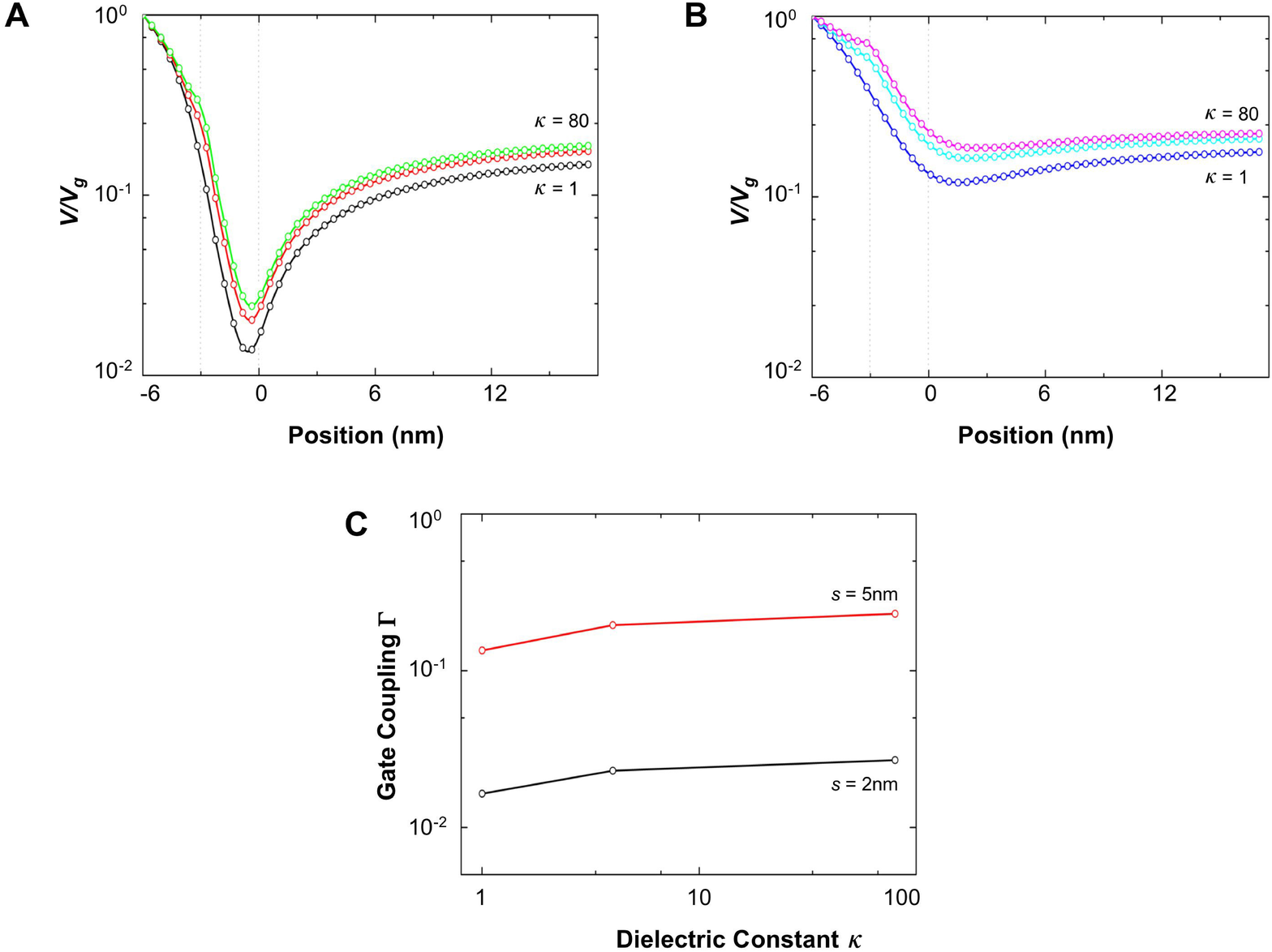}
\caption{(Color Online) {\bf Electrodes with non-uniform cross-section -- role of the dielectric constant. {\bf (A)}} Normalized potential for the three-dimensional device shown in Figure 4A with $L_{ox}=3$nm, $t=3$nm, $s=2$nm, $\theta=120^\circ$, values of $\kappa$ from bottom to top are 1, 4, and 80; {\bf (B)} Normalized potential for the three-dimensional device shown in Figure 4A with $L_{ox}=3$nm, $t=3$nm, $s=5$nm, $\theta=120^\circ$, values of $\kappa$ from bottom to top are 1, 4, and 80; {\bf (C)} Gate coupling for the three-dimensional device shown in Figure 4A with $L_{ox}=3$nm, $t=3$nm, $\theta=120^\circ$.}
\end{flushright}
\end{figure}
\newpage

\begin{figure}
\begin{flushright}
\includegraphics[width=5.5in]{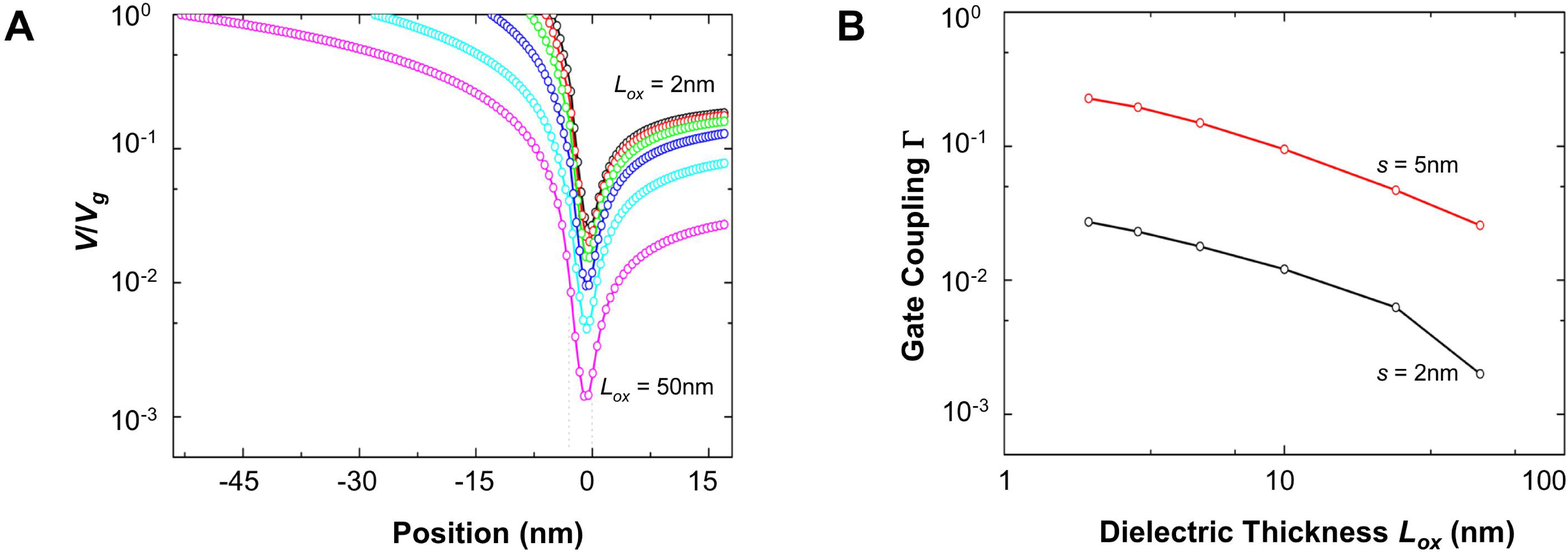}
\caption{(Color Online) {\bf Electrodes with non-uniform cross-section -- role of the dielectric thickness. {\bf (A)}} Normalized potential for the three-dimensional device shown in Figure 4A with $t=3$nm, $s=2$nm, $\theta=120^\circ$, $\kappa=4$, values of $L_{ox}$ from top to bottom are 2, 3, 5, 10, 25, and 50nm; {\bf (B)} Gate coupling as a function of dielectric thickness for the three-dimensional device shown in Figure 4A with $t=3$nm, $\theta=120^\circ$, $\kappa=4$.}
\end{flushright}
\end{figure}
\newpage

\begin{figure}
\begin{flushright}
\includegraphics[width=5.5in]{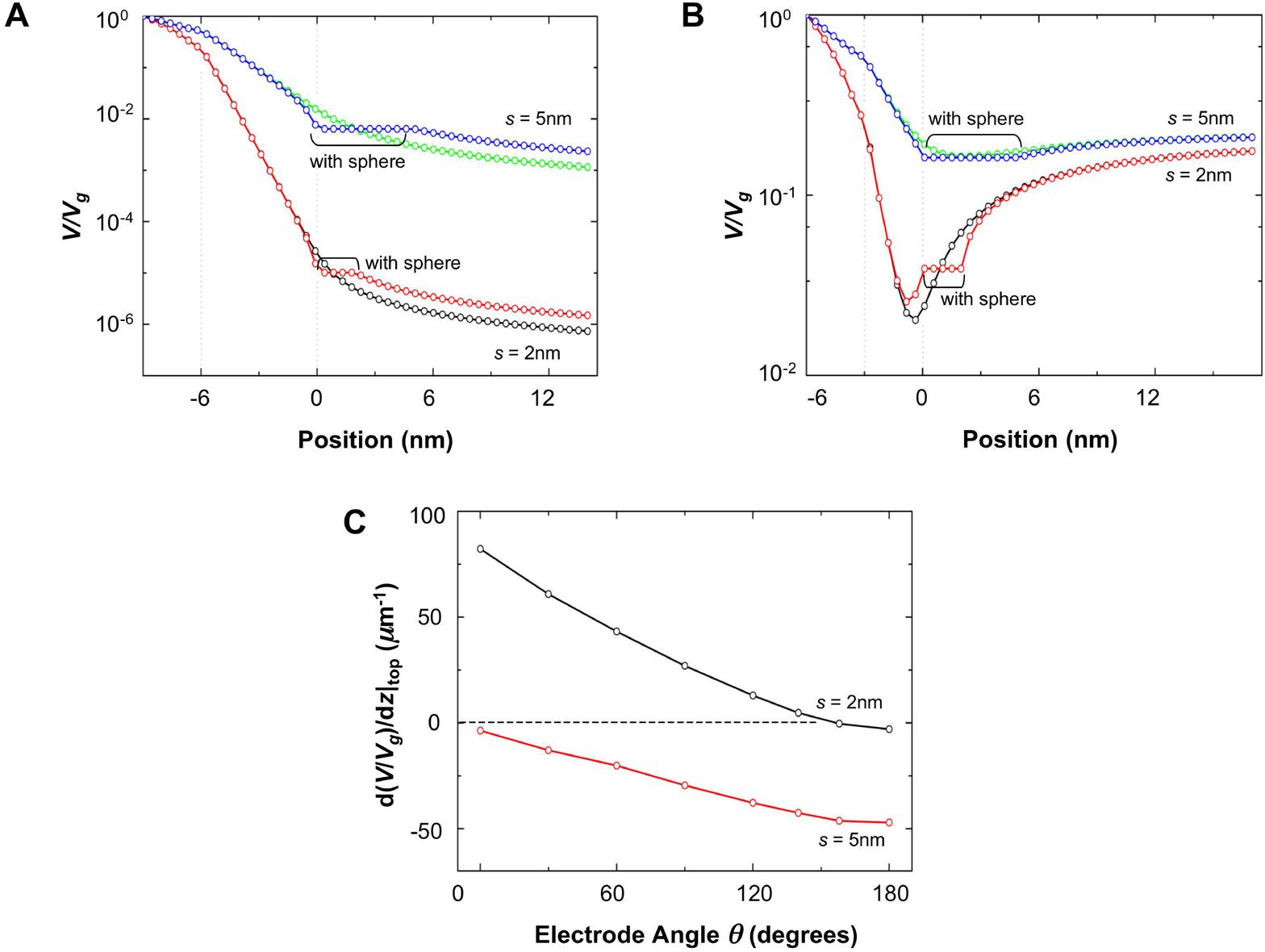}
\caption{(Color Online) {\bf Electrodes with uniform and non-uniform cross-section -- effect of molecular polarizability. {\bf (A)}} Normalized potential for the two-dimensional device shown in Figure 1A with $L_{ox}=3$nm, $t=6$nm, $\kappa=4$, for bare electrodes and electrodes with a polarizable sphere of diameter $s$ and $\kappa=10^{6}$ positioned at the top of the nanogap; {\bf (B)} Normalized potential for the three-dimensional device shown in Figure 4A with $L_{ox}=3$nm, $t=3$nm, $\kappa=4$, for bare electrodes and electrodes with a polarizable sphere of diameter $s$ and $\kappa=10^{6}$ positioned at the top of the nanogap; {\bf (C)} Derivative of the normalized potential with respect to vertical position for the three-dimensional device shown in Figure 4A with $L_{ox}=3$nm, $t=3$nm, $\kappa=4$.}
\end{flushright}
\end{figure}
\newpage

	\end{document}